\documentclass[prb,superscriptaddress,twocolumn]{revtex4}

\usepackage{epsfig,amsmath}

\renewcommand{\vec}[1]{{\rm\bf #1}}

\newcommand{\ep}{\epsilon}
\newcommand{\unitmatrix}{\openone}

\begin{document}

\title{Interplay of Coulomb and electron-phonon interactions in graphene}
\author{D.~M.~Basko}\email{basko@sissa.it}
\affiliation{International School of Advanced Studies (SISSA), via
Beirut~2-4, 34014 Trieste, Italy}
\author{I.~L.~Aleiner}
\affiliation{Physics Department, Columbia University, New York, NY
10027, USA}

\begin{abstract}
We consider mutual effect of the electron-phonon and strong
Coulomb interactions on each other by summing up leading
logarithmic corrections via the renormalization group approach. We
find that the Coulomb interaction enhances electron coupling to
the intervalley $A_1$~optical phonons, but not to the intravalley
$E_2$~phonons.
\end{abstract}

\maketitle

Electron-phonon coupling (EPC) in graphene
is currently a subject of intense research. Experimental
information on EPC is obtained by Raman
spectroscopy\cite{Ferrari,Gupta,Graf,Pisana,Jun} and
angle-resolved photoemission spectroscopy
(ARPES).\cite{Rotenberg} Theoretically, EPC constants are usually
calculated from density-functional theory
(DFT),\cite{Lazzeri,Mauri,Cohen}
where the
exchange (Fock) term is treated in the local density approximation
(LDA),\cite{LDA} or the generalized gradient approximation
(GGA).\cite{GGA}
Disagreement between the calculated EPC and the ARPES data has
been pointed out in Ref.~\onlinecite{Mauri}.

Also, the ratio between calculated EPC constants for different
phonon modes disagrees with the ratio of the integrated intensities
$I_{D^*}/I_{G^*}$ of the corresponding two-phonon Raman peaks, as
noted in Ref.~\onlinecite{myself}. Namely, in the Raman spectrum of
graphene two two-phonon peaks are seen: the so-called $D^*$ peak
near $2\omega_{A_1}=2650\:\mathrm{cm}^{-1}$, and the $G^*$ peak
near $2\omega_{E_2}=3250\:\mathrm{cm}^{-1}$, corresponding to
scalar $A_1$~phonons from the vicinity of the $K$~point of the
first Brillouin zone, and to pseudovector $E_2$~phonons from the
vicinity of the $\Gamma$~point, respectively. Experimentally,
$I_{D^*}/I_{G^*}\approx{20}$,\cite{Ferrari} which cannot be
reproduced with the EPC constants obtained from DFT calculations.

In this work we consider mutual effect of weak electron-phonon and
strong Coulomb interactions on each other by summing up leading
logarithmic corrections via the renormalization group (RG)
approach, which goes beyond Hartree-Fock approximation. Coulomb
interaction has been known to be a source of logarithmic
renormalizations for Dirac fermions.\cite{AbrikosovBeneslavskii,Guinea94,Guinea99}
Coulomb renormalizations in
graphene subject to a magnetic field have been considered in
Ref.~\onlinecite{AleinerTsvelik}, Coulomb effect on static disorder has
been studied in Refs.~\onlinecite{Ye,Guinea2005,AleinerFoster}. To the
best of our knowledge, Coulomb renormalization of EPC has never
been considered; moreover, at energies higher than the
phonon frequency, EPC itself is a source of logarithmic
renormalizations, and has to be included in the RG procedure.

Upon solution of the RG equations we obtain that (i)~EPC tends to
enhance Coulomb interaction, but not sufficiently to dominate over
the flow to weak coupling, found earlier;\cite{Guinea99}
(ii)~Coulomb interaction enhances the EPC only for the scalar
$A_1$~phonons, while renormalization of the coupling to the
pseudovector $E_2$~phonons is only due to EPC and is relatively
weak, in agreement with the Raman data.\cite{Ferrari}

\begin{table}
\begin{tabular}[t]{|c|c|c|c|c|c|c|} \hline
$C_{6v}$ & $E$ & $C_2$ & $2C_3$ & $2C_6$ & $\sigma_{a,b,c}$ &
$\sigma_{a,b,c}'$
\\ \hline\hline $A_1$ & 1 & 1 & 1 & 1 & 1 & 1 \\ \hline $A_2$ & 1
& 1 & 1 & 1 & $-1$ & $-1$ \\ \hline $B_2$ & 1 & $-1$ & 1 & $-1$ &
$1$ & $-1$ \\ \hline $B_1$ & 1 & $-1$ & 1 & $-1$ & $-1$ & $1$ \\
\hline $E_1$ & 2 & $-2$ & $-1$ & $1$ & 0 & 0 \\ \hline $E_2$ & 2 &
2 & $-1$ & $-1$ & 0 & 0 \\ \hline\end{tabular}\hspace{1cm}
\caption{Irreducible representations of the group $C_{6v}$ and
their characters.\label{tab:C6vC3v}}
\end{table}

\begin{table}
\begin{tabular}{|c|c|c|c|c|c|c|} \hline
irrep & 
$A_1$ & $B_1$ & $A_2$ & $B_2$ & $E_1$ & $E_2$ \\ \hline
\multicolumn{7}{|c|}{valley-diagonal matrices}\\
\hline
matrix & $\unitmatrix$ & $\Lambda_z$ & $\Sigma_z$ &
$\Lambda_z\Sigma_z$ & $\Sigma_x,\,\Sigma_y$ &
$-\Lambda_z\Sigma_y,\Lambda_z\Sigma_x$ \\ \hline
\multicolumn{7}{|c|}{valley-off-diagonal matrices} \\
\multicolumn{7}{|c|}{$\qquad\quad\;
\overbrace{\qquad\qquad\quad\;}\;
\overbrace{\qquad\qquad}\;
\overbrace{\qquad\qquad\qquad\qquad\qquad\qquad\;}$} \\
\hline
matrix & $\Lambda_x\Sigma_z$ &
$\Lambda_y\Sigma_z$ & $\Lambda_x$ & $\Lambda_y$ &
$\Lambda_x\Sigma_y,-\Lambda_x\Sigma_x$ &
$\Lambda_y\Sigma_x,\Lambda_y\Sigma_y$ \\ \hline
\end{tabular}
\caption{Classification of $4\times{4}$ hermitian matrices
by irreducible representations of the $C_{6v}$~group.
Matrices joined by braces transform through each
other under translations.\label{tab:matrices}}
\end{table}

We measure the single-electron energies from the
Fermi level of the undoped (half-filled) graphene. The Fermi
surface of undoped graphene consists of two points, called
$K$~and~$K'$. Graphene unit cell contains two atoms, each of them
has one $\pi$-orbital, so there are two electronic states for each
point of the first Brillouin zone (we disregard the electron
spin). Thus, there are exactly four electronic states with zero
energy. An arbitrary linear combination of them is represented by
a 4-component column vector~$\psi$. States with low energy are
obtained by including a smooth position dependence
$\psi(\vec{r})$, $\vec{r}\equiv(x,y)$. The low-energy hamiltonian
has the Dirac form:~\cite{Wallace}
\begin{equation}\label{Hel=}
\hat{H}_{el}=\int{d}^2\vec{r}\,\hat\psi^\dagger(\vec{r})\,
(-iv\vec\Sigma\cdot\vec\nabla)\,\hat\psi(\vec{r}).
\end{equation}
We prefer not to give the explicit form of the isospin matrices
$\vec\Sigma\equiv(\Sigma_x,\Sigma_y)$, which depends on the choice
of basis (specific arrangement of the components in the
column~$\psi$). We only note that all 16 generators of the $SU(4)$
group, forming the basis in the space of $4\times{4}$ hermitian
matrices, can be classified according to the irreducible
representations of~$C_{6v}$, the point group of the graphene
crystal (Tables \ref{tab:C6vC3v} and~\ref{tab:matrices}). They can
be represented as products of two mutually commuting algebras of
Pauli matrices $\Sigma_x,\Sigma_y,\Sigma_z$ and
$\Lambda_x,\Lambda_y,\Lambda_z$,\cite{Falko,AleinerEfetov} which
fixes their algebraic relations. By definition,
$\Sigma_x,\Sigma_y$ are the matrices, diagonal in the $K,K'$
subspace, and transforming according to the $E_1$~representation
of~$C_{6v}$. The Fermi velocity $v\approx{10}^8\:\mbox{cm/s}$.

The hamiltonian of the long-range Coulomb interaction between
electrons has the form (hereinafter we imply the summation over
the spin indices):
\begin{equation}
\hat{H}_{ee}=\frac{e^2}2\int{d}^2\vec{r}\,d^2\vec{r}'\,
\frac{\hat\rho(\vec{r})\hat\rho(\vec{r}')}{|\vec{r}-\vec{r}'|},\quad
\hat\rho(\vec{r})=
\hat\psi^\dagger(\vec{r})\hat\psi(\vec{r}).
\end{equation}
The background
dielectric constant of the substrate is incorporated into~$e^2$.

\begin{figure}
\includegraphics[width=8cm]{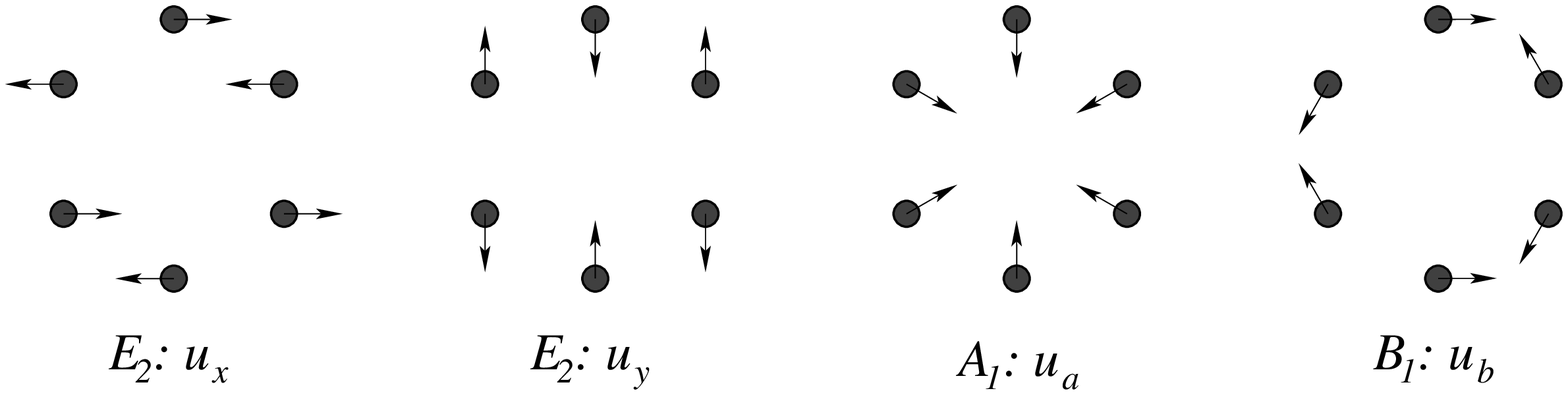}
\caption{\label{fig:phonons} Phonon displacements for $E_2$, $A_1$,
  and $B_1$ modes.}
\end{figure}

For low-energy electronic states EPC is efficient if the phonon
wave vector is close to $\Gamma$,~$K$ or $K'$ point. Considering
only in-plane displacements, we have 4 degrees of freedom per unit
cell. Consider the $\Gamma$~point first. Two modes are acoustic,
they weakly couple to electrons, and are neglected. The other two
correspond to $E_2$ (pseudovector) optical phonons, shown in
Fig.~\ref{fig:phonons}. They couple to the electronic motion via
the $KK'$-diagonal $E_2$ matrices from Table~\ref{tab:matrices}.
The $K$ and $K'$ points are related by the time reversal symmetry,
so the phonon frequencies are the same, and one can form real
linear combinations of the modes from $K$ and~$K'$. They transform
according to $A_1,B_1,A_2,B_2,E_1,E_2$ representations of~$C_{6v}$,
and couple to the electronic motion via corresponding
$KK'$-off-diagonal matrices.
Linear coupling to $A_2$~and~$B_2$ displacements is forbidden by
time-reversal symmetry, and coupling to the $E_1,E_2$ modes is
numerically small.\cite{Piscanec} The reason for this smallness
is that $E_1,E_2$ displacements do not change any C--C bond
length; in the tight-binding approximation this coupling simply
vanishes. Thus, we restrict our attention to the $E_2$ modes from
the $\Gamma$~point and $A_1$~and~$B_1$ combinations of the modes
from the $K,K'$ points, shown in Fig.~\ref{fig:phonons}. They
are the only modes seen in the Raman spectra of
graphene.\cite{Ferrari,Gupta,Graf,Pisana,Jun}

We take the magnitude of the carbon atom displacement as the
normal coordinate for each mode, denoted by $u_\mu$, $\mu=x,y,a,b$
for the four modes, shown in Fig.~\ref{fig:phonons}, respectively.
Upon quantization of the phonon field, $\hat{u}_\mu$ and the
phonon hamiltonian $\hat{H}_{\mathrm{ph}}$ are expressed in terms
of the creation and annihilation operators
$\hat{b}^\dagger_{\vec{q}\mu},\hat{b}_{\vec{q}\mu}$, as
\begin{equation}
\hat{u}_\mu(\vec{r})=
\sum_{\vec{q}}
\frac{\hat{b}_{\vec{q}\mu}e^{i\vec{q}\vec{r}}
+\mathrm{h.c.} } {\sqrt{2NM\omega_\mu}},\quad
\hat{H}_{ph}=\sum_{\vec{q},\mu}
\omega_\mu\hat{b}_{\vec{q}\mu}^\dagger\hat{b}_{\vec{q}\mu}.
\end{equation}
The crystal is assumed to have the area $L_xL_y$, and to contain
$N$~carbon atoms of mass~$M$. The $\vec{q}$~summation is performed
as $\sum_\vec{q}\to{L}_xL_y\int{d}^2\vec{q}/(2\pi)^2$. ``h.c.''
stands for hermitian conjugate. The two degenerate $E_2$~modes
have the frequency $\omega_{E_2}\approx{0}.196\:\mbox{eV}$.
As the $A_1$~and~$B_1$ modes represent real linear combinations
of modes from $K,K'$ points, they have the same frequency,
$\omega_{A_1}\approx{0}.170\:\mbox{eV}$, and appear with the same
coupling constant in the EPC hamiltonian:
\begin{eqnarray}
&&\hat{H}_{EPC}=\int{d}^2\vec{r}\,\hat\psi^\dagger(\vec{r})
\left[\sum_{\mu}F_\mu\hat{u}_\mu(\vec{r})(\Lambda\Sigma)_\mu
\right]\hat\psi(\vec{r})=\nonumber\\
&&=\int{d}^2\vec{r}\,\hat\psi^\dagger(\vec{r})
\left\{F_{E_2}\left[\hat{u}_x(\vec{r})\Lambda_z\Sigma_y
-\hat{u}_y(\vec{r})\Lambda_z\Sigma_x\right]\right.+\nonumber\\
&&\qquad\quad
+\left.F_{A_1}\left[\hat{u}_a(\vec{r})\Lambda_x\Sigma_z
+\hat{u}_b(\vec{r})\Lambda_y\Sigma_z\right]\right\}
\hat\psi(\vec{r}).\label{Heph=}
\end{eqnarray}
The coupling constants $F_{E_2}$ and $F_{A_1}$ are not related by
any symmetry. However, in the tight-binding model
$F_{E_2}=F_{A_1}=3(\partial{t}_0/\partial{a})$, where $t_0$~is the
nearest-neighbor coupling matrix element, and $a$~is the bond
length.

\begin{figure}
\includegraphics[width=8cm]{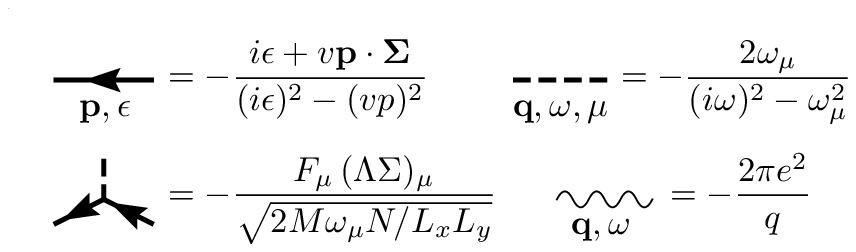}
\caption{\label{fig:diagrams} Analytical expressions of the
diagrammatic technique.}
\end{figure}

\begin{figure}
\includegraphics[width=8cm]{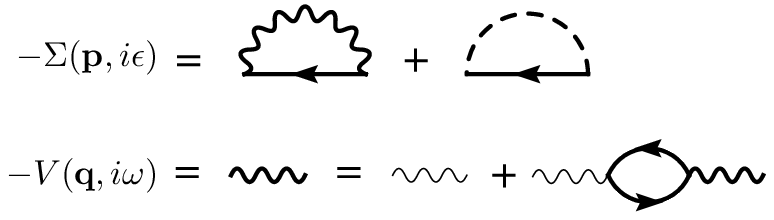}
\caption{\label{fig:selfenrg} Electron self-energy due to the
screened Coulomb interaction and the EPC.}
\end{figure}

As we are interested in
energies much higher than the temperature~$T$, we set $T=0$;
still, calculations are much more transparent in the Matsubara
representation. Electron and phonon Green's functions are given by
\begin{equation}
G(\vec{p},i\ep)=-\frac{i\ep+v\vec{p}\vec\Sigma}{\ep^2+(vp)^2},\quad
D_\mu(i\omega)=-\frac{2\omega_\mu}{\omega^2+\omega_\mu^2},
\end{equation}
their graphical representation is shown in Fig.~\ref{fig:diagrams}.
The electronic self-energy is a sum of the Coulomb\cite{Guinea99}
and EPC\cite{Guinea81} contributions:
$\Sigma=\Sigma^{ee}+\Sigma^{ph}$, shown in
Fig.~\ref{fig:selfenrg}.

The leading logarithmic term in $\Sigma^{ee}$ is given by:\cite{Guinea99}
\begin{eqnarray}
&&\Sigma^{ee}(\vec{p},i\ep)=
-\int\frac{d\omega}{2\pi}\frac{d^2\vec{q}}{(2\pi)^2}\,
V(\vec{q},i\omega)\,G(\vec{p}-\vec{q},i\ep-i\omega)\nonumber\\
&&\approx\frac{8}{\pi^2\mathcal{N}}\left[f(g)(2i\ep-v\vec{p}\vec\Sigma)
-\tilde{f}(g)(i\ep-v\vec{p}\vec\Sigma)\right]
\ln\frac{\xi_{max}}{\xi_{min}},\nonumber\\ &&\label{SigmaCoul=}\\
&&f(g)=1-\frac{\pi}{2g}+\frac{\arccos{g}}{g\sqrt{1-g^2}},\;\;\;
\tilde{f}(g)=\frac{g\arccos{g}}{\sqrt{1-g^2}},\\
&&V(\vec{q},i\omega)= \frac{16g}{\mathcal{N}}\frac{v}q\,
\frac{\sqrt{(vq)^2+\omega^2}}{gvq+\sqrt{(vq)^2+\omega^2}}, \;\;\;
g=\frac{\pi\mathcal{N}e^2}{8v}. \label{VRPA=}
\end{eqnarray}
$\mathcal{N}=4$ is the number of Dirac species, valley and spin
degeneracy taken into account.
The lower cutoff $\xi_{min}\sim\max\{vp,\ep\}$, the upper cutoff
$\xi_{max}\sim{v}/a$ is of the order of the electronic bandwidth.
The logarithmic divergence in the Fock self-energy~$\Sigma^{ee}$
is due to long-distance nature of the Coulomb interaction, and thus
is not picked up by local approximations such as LDA or GGA.
The random phase approximation (RPA) for $V(\vec{q},i\omega)$,
shown in Fig.~\ref{fig:selfenrg}, corresponds to expansion of the
pre-logarithm coefficient to the leading order in the parameter
$1/\mathcal{N}=0.25$, assumed to be small. This is justified
better than expansion in the dimensionless coupling constant~$g$,
obtained with the bare coupling $2\pi{e}^2/q$.
Indeed, for $\mathcal{N}=4$ we have
$g=(\pi/2)(e^2/v)\approx{3}.4$; background
dielectric screening reduces it to $g\sim{1}$. 

The leading logarithmic asymptotics of $\Sigma^{ph}$ is given by
\begin{eqnarray}
&&\Sigma^{ph}(i\ep)=
-\int\frac{d\omega}{2\pi}\frac{d^2\vec{q}}{(2\pi)^2}\sum_\mu
\frac{F^2_\mu}{2M\omega_\mu}\frac{\sqrt{27}a^2}4\,
D_\mu(i\omega)\times\nonumber\\ &&\qquad\qquad\qquad{}\times
(\Lambda\Sigma)_\mu{G}(\vec{p}-\vec{q},i\ep-i\omega)(\Lambda\Sigma)_\mu
\approx\nonumber\\
&&\approx\frac{\lambda_{E_2}+\lambda_{A_1}}{{2\pi}}\:i\ep
\ln\frac{\xi_{max}}{\xi_{min}},\label{SigmaEPC=}\quad
\lambda_\mu=\frac{F^2_\mu}{M\omega_\mu{v}^2}\frac{\sqrt{27}a^2}4.
\end{eqnarray}
Here $\xi_{min}\sim\max\{\ep,\omega_\mu\}$, $\xi_{max}\sim{v}/a$,
and $\sqrt{27}a^2/4$ is the area per carbon atom. The
dimensionless constants $\lambda_{E_2},\lambda_{A_1}$ will be
treated as small parameters.

The latter statement deserves some discussion. In principle, one
could proceed analogously to the Coulomb case: instead of doing
the perturbative expansion in $\lambda_\mu$, one could dress the
bare phonon propagators by the appropriate polarization operators
$\Pi(\vec{q},i\omega)$, corresponding to $1/\mathcal{N}$
expansion. Since $\Pi(\vec{q},i\omega)\propto{q}$ at
$\omega\ll{q}\ll{1}/a$,\cite{Shung}
the dressed phonon frequency would grow as~$\sqrt{q}$, and
$\Sigma^{ph}$ would no longer diverge logarithmically. However,
the inelastic X-ray scattering data for the phonon
dispersion~\cite{Maultzschexp} show that the phonon dispersion is
smaller than the phonon frequency itself. Thus, the
renormalization of the phonon frequency remains small even at
$q\sim{1}/a$, so the perturbative expansion in~$\lambda_\mu$ is
more justified.

The logarithmically divergent integrals in Eqs.~(\ref{SigmaCoul=})
and~(\ref{SigmaEPC=}) have different structure due to different
form of the screened interaction $V(\vec{q},i\omega)$ and the
phonon propagator $D_\mu(i\omega)$. In Eq.~(\ref{SigmaCoul=}) the
integral is dominated by the frequencies $|\omega|\sim{v}q$, while
in Eq.~(\ref{SigmaEPC=}) it is $|\omega|\sim\omega_\mu$, since
$D_\mu(i\omega)\propto{1}/\omega^2$ at $|\omega|\gg\omega_\mu$.
Thus, in the calculation of the leading logarithmic asymptotics it
is sufficient to approximate
$D_\mu(i\omega)\approx -2\pi\delta(\omega)$.
This substitution makes the phonon propagator (combined with EPC
vertices) formally analogous to the correlator of static disorder
potential (i.~e., from the point of view of electrons with
$\ep\gg\omega_\mu$ the lattice is effectively frozen).
Thus, renormalizations due to EPC at $\ep\gg\omega_\mu$ are
equivalent to those due to static
disorder~\cite{Ye,Guinea2005,AleinerEfetov,AleinerFoster}. This
equivalence holds only in the leading order in EPC, since in
higher orders the phonon propagator is dressed by polarization
loops, and the static disorder correlator is not.

The presence of the large logarithm invalidates the first-order expansion
in $1/\mathcal{N}$, and makes it necessary to sum all leading
logarithmic terms $\sim(1/\mathcal{N})^n\ln^n$ of the perturbation
theory. This is done using the standard RG
procedure.\cite{AbrikosovBeneslavskii,%
Guinea94,Guinea99,AleinerTsvelik,Ye,Guinea2005,AleinerFoster}
Let us introduce the running cutoff $\xi_{max}e^{-\ell}$. One RG
step consists of reducing the cutoff, $\ell\to\ell+\delta\ell$, so that
$e^{-\delta\ell}\ll{1}$, while
$(1/\mathcal{N})\delta{\ell}\ll{1}$,
$\lambda_\mu\delta\ell\ll{1}$. The inverse Green's function
transforms as
\begin{equation}
i\ep-v\vec{p}\vec\Sigma-\Sigma(\vec{p},i\ep)=
\frac{i\ep-(v+\delta{v})\vec{p}\vec\Sigma}{1+\delta{Z}}\,,
\end{equation}
where $\delta{Z}$ is chosen to preserve the coefficient at $i\ep$ upon
rescaling of the electronic fields,
$\psi\to(1+\delta{Z}/2)\psi$:
$\delta{Z}=\partial\Sigma(\vec{p},i\ep)/\partial(i\ep)$. Then
$v$~is renormalized as
\begin{equation}
\frac{\delta{v}}{v}=\frac{\partial\Sigma(\vec{p},i\ep)}{\partial(i\ep)}
+\frac{\partial\Sigma(\vec{p},i\ep)}{\partial(v\vec{p}\vec\Sigma)}.
\end{equation}


\begin{figure}
\includegraphics[width=8cm]{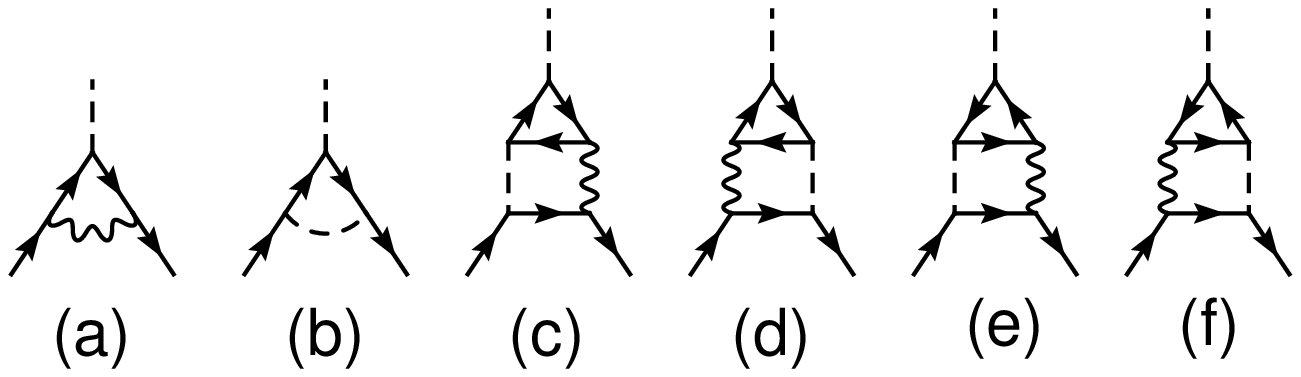}
\caption{\label{fig:PhononVertexRG} Logarithmic corrections
  to the EPC vertex $F_\mu$ of the order
  $O(1/\mathcal{N},\lambda_\mu^2)$. Diagrams (c)--(f) vanish.}
\end{figure}

\begin{figure}
\includegraphics[width=3cm]{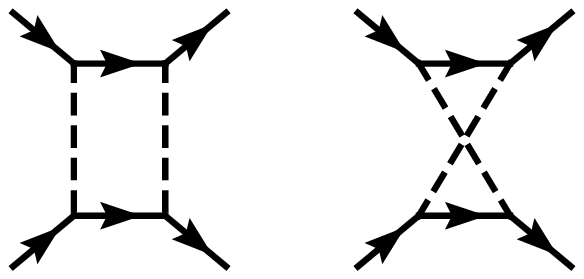}
\caption{\label{fig:Phonon2LineRG} Logarithmic diagrams of the
order $O(\lambda_\mu^2)$ not reduced to a renormalization of the
EPC vertex~$F_\mu$.}
\end{figure}

Next, we determine renormalization of the coupling constants. The
electron charge~$e$ is not changed, as guaranteed by the gauge
invariance, so the renormalization of the Coulomb coupling
constant~$g$ is determined by the velocity~$v$. For the EPC,
logarithmic vertex corrections of the order
$O(1/\mathcal{N},\lambda_\mu^2)$ are shown in
Fig.~\ref{fig:PhononVertexRG}. Two other diagrams
(Fig.~\ref{fig:Phonon2LineRG}) should be taken into account,
as they are of the same order and also logarithmic.
As a result, we obtain the following RG equations:
\begin{subequations}
\begin{eqnarray}
\frac{dg}{d\ell}&=&-\frac{8f(g)}{\pi^2\mathcal{N}}\,g
+\frac{\lambda_{E_2}+\lambda_{A_1}}{2\pi}\,g,\label{RGflowg=}\\
\frac{d\lambda_{E_2}}{d\ell}&=&\frac{\lambda_{A_1}^2}{2\pi},\\
\frac{d\lambda_{A_1}}{d\ell}&=&
\frac{16f(g)}{\pi^2\mathcal{N}}\,\lambda_{A_1}.\label{RGflowwA1=}
\end{eqnarray}\label{RGflow=}
\end{subequations}
Because of the diagrams of Fig.~\ref{fig:Phonon2LineRG},
the renormalized~$\lambda_\mu$ cannot be related to a new EPC
vertex~$F_\mu$. Iterations of these diagrams generate
electron coupling to multiphonon excitations, all included in the
renormalized~$\lambda_\mu$.

The coupling constant~$\lambda_{E_2}$ at energies
$\sim\omega_{E_2}\approx{0}.2\:\mbox{eV}$ can be extracted from
the experimental data of Ref.~\onlinecite{Jun} (change of the Raman
$G$~peak with the electron density):
$\lambda_{E_2}\approx{0}.035$, corresponding to
$F_{E_2}=6\:\mathrm{eV}/\mbox{\AA}$. Then the main effect comes
from the Coulomb terms: 
as $f(g)>0$, $g$~flows to weak coupling,\cite{Guinea99}
$\lambda_{A_1}$~is enhanced, while the enhancement
of~$\lambda_{E_2}$, proportional to $\lambda_{A_1}^2$, is much
weaker due to the cancellation between Coulomb self-energy and
vertex corrections. This behavior in qualitative agreement with
the Raman data: when the $\lambda_\mu^2$ term is neglected, the
ratio of the intensities of the two-phonon peaks, mentioned in the
introduction, is
$I_{D^*}/I_{G^*}=2(\lambda_{A_1}/\lambda_{E_2})^2$.\cite{myself}
If only Coulomb terms are kept in Eqs. (\ref{RGflow=}), their
integration gives
$\lambda_{A_1}(\ell)/\lambda_{A_1}(0)=[g(0)/g(\ell)]^2=
[v(\ell)/v(0)]^2$, which, in principle, can be checked experimentally.

\begin{figure}
\includegraphics[width=8cm]{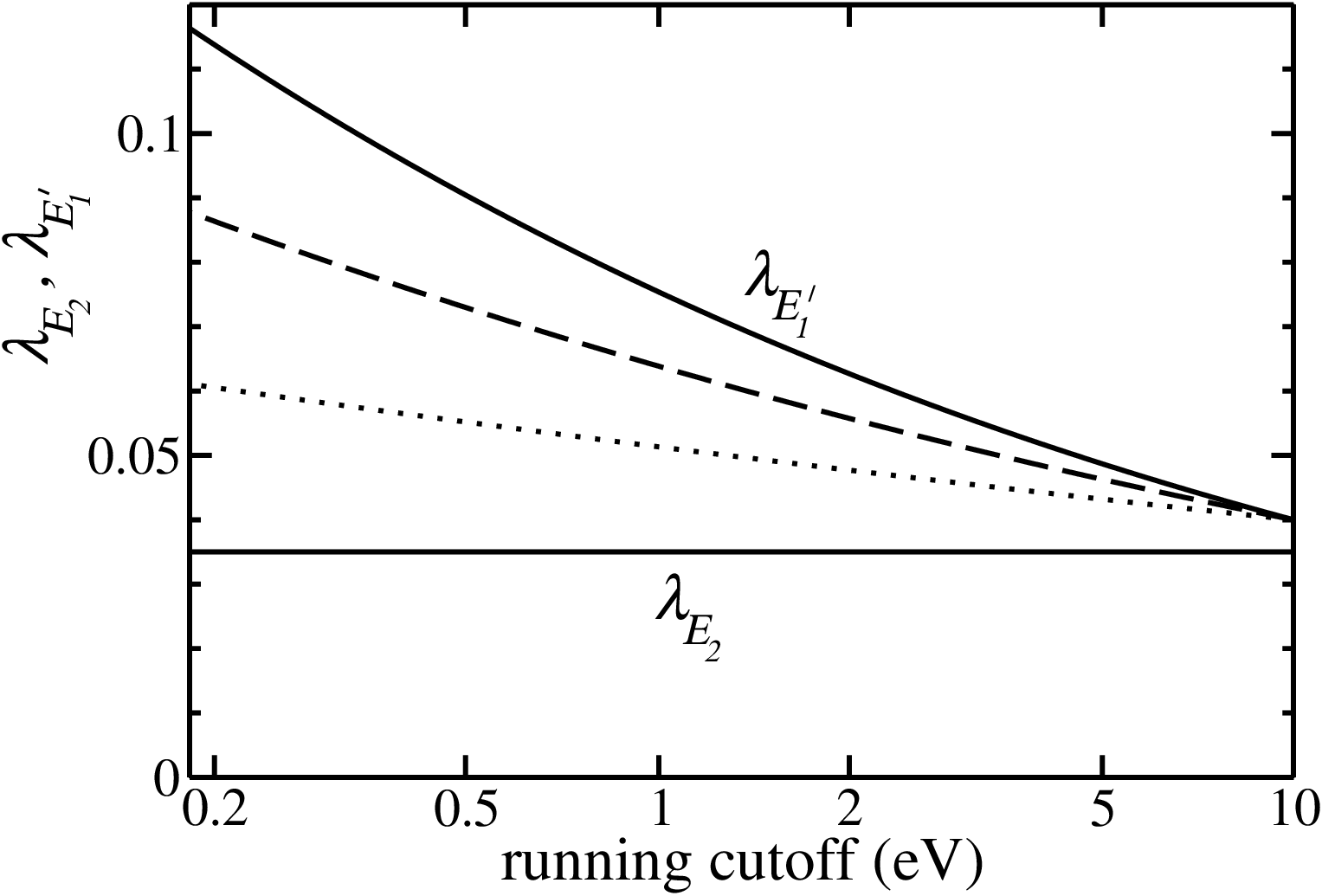}
\caption{\label{fig:RGflow} Flow of the dimensionless coupling
constants $\lambda_{A_1}$ (three upper curves, starting from the
bare value 0.04 at 10~eV, respectively) for three values of the
bare Coulomb coupling $g(0)=3.4,\:1.5,\:0.5$ (solid, dashed, and
dotted curves). The constant
$\lambda_{E_2}=0.035$ is unchanged in the order $O(\lambda)$.}
\end{figure}

To study the behavior of the coupling constants quantitatively, we
solve Eqs.~(\ref{RGflow=}) numerically, neglecting
the~$\lambda_\mu^2$ term. The largest value of~$\ell$ is determined
by the lower cutoff
$\xi_{min}\sim\omega_\mu\sim{0}.2\:\mbox{eV}$.  In
Fig.~\ref{fig:RGflow}, we show the flow of~$\lambda_{A_1}$ for
three values of the bare Coulomb coupling constant: $g(0)=3.4$
(corresponding to no dielectric screening at all), $g(0)=1.5$, and
$g(0)=0.5$. The bare values of the the electron-phonon coupling
constants $\lambda_{E_2}(0)=0.035$, $\lambda_{A_1}(0)=0.040$ were
chosen (a)~to satisfy the relation
$\lambda_{E_2}(0)/\lambda_{A_1}(0)=\omega_{A_1}/\omega_{E_2}$,
valid in the tight-binding approximation, (b)~to reproduce the
experimental value $\lambda_{E_2}\approx{0}.035$. Note that the RPA
calculation without the RG collection of all leading logarithmic terms,
would give all dependencies on Fig.~\ref{fig:RGflow} to be straight
lines with slopes fixed at 10~eV. A comparable error would be
produced by the GW approximation, which neglects vertex corrections,
and thus picks up correctly only the first term of the logarithmic series.

To estimate the EPC strength relevant for Raman scattering, we stop
the RG flow at electronic energies $\sim{1}\:\mbox{eV}$ (half of the
incident laser frequency).
In the unscreened case, $g(0)=3.4$, it gives
$\lambda_{A_1}/\lambda_{E_2}\approx{3.2}$, in agreement with the
observed ratio $I_{D^*}/I_{G^*}\approx{20}$.

To conclude, in this paper we have considered mutual
effect of the weak electron-phonon and strong Coulomb interactions
on each other by summing up leading logarithmic corrections via
the renormalization group approach in the intermediate energy
range $\omega_{E_2},\omega_{A_1}<\ep<v/a$. At these energies
quantum fluctuations of the phonon field may be viewed as
effective static disorder. We find that Coulomb interaction
enhances electron coupling to the intervalley $A_1$~optical
phonons, but not to the intavalley $E_2$~phonons, in agreement
with the experimental data for two-phonon Raman scattering.

We thank M.~S.~Foster, F.~Guinea, and F.~Mauri for helpful
discussions.

\end{document}